\documentclass[letter,twocolumn]{jpsj3}
\usepackage{txfonts}

\title{Diffusion Thermopower of Quantum Hall States Measured in Corbino Geometry}

\author{\name{Shuhei \surname{Kobayakawa}}, \name{Akira \surname{Endo}}\thanks{E-mail address: akrendo@issp.u-tokyo.ac.jp}, and \name{Yasuhiro \surname{Iye}} 
}
\inst{The Institute for Solid State Physics, The University of Tokyo, \address{5-1-5 Kashiwanoha, Kashiwa, Chiba 277-8581, Japan} 
}
\abst{We have measured the diffusion thermopower of a quantum Hall system in a Corbino setup. A concentric electron-temperature gradient is introduced by irradiating  microwaves, via a coplanar waveguide, near the outer rim of a circular mesa of a two-dimensional electron gas. The resulting radial thermovoltages exhibit sawtooth-like oscillations with the magnetic field, taking large positive (negative) values just below (above) integer fillings with sign reversal at the center of the quantum Hall plateaus. The behavior is in agreement with a recent theory [Y.\ Barlas and K.\ Yang: Phys.\ Rev.\ B \textbf{85} (2012) 195107], which treats disorder within the self-consistent Born approximation.}

\kword{diffusion thermopower, quantum Hall effect, Corbino disk, microwave heating, coplanar waveguide, conductivity, two-dimensional electron gas}

\begin{document}
\maketitle

The thermopower of quantum Hall (QH) systems has been attracting interest as a sensitive probe to examine the electronic properties of the systems \cite{Jonson84,Obloh86,Fletcher86,Ying94,Fletcher99,Zalinge03,Granger09,Bergman10}, which is further spurred by a recent proposal of the possibility to explore intriguing statistics of the quasiparticles in the $\nu = 5/2$ fractional QH state \cite{Yang09,Barlas12,Chickering10,Chickering13}. The vast majority of the measurements have been performed in the Hall bar geometry thus far (see, e.g., ref.\ \citen{Fletcher99} and references therein). As will be detailed below, the radial thermopower $S_{rr}$ measured in the Corbino geometry is qualitatively different from the longitudinal thermopower $S_{xx}$ measured in the Hall bar geometry \cite{Zalinge03,Barlas12}.
The purpose of the present paper is to report our measurements of the diffusion thermopower in the Corbino geometry for integer QH states.

The thermopower generally contains contributions from two distinct mechanisms: diffusion and phonon drag \cite{Fletcher99}. The high sensitivity to the electronic properties is expected for the former contribution. In a two-dimensional electron gas (2DEG) embedded in a GaAs/AlGaAs heterostructure, however, it is well known that the latter dominates the measured thermopower \cite{Fletcher99,Zalinge03}, unless measurements are done at very low temperatures $\lesssim 150$ mK \cite{Ying94,Chickering10}; in standard measurements using an external heater to introduce the temperature gradient $\boldsymbol{\nabla} T$, the resulting heat current is predominantly carried by the phonons in the host crystal, leading to the dominance of the phonon-drag contribution despite the weakness of the electron-phonon interaction. The measurement of Corbino thermopower in QH systems at 1.5 K was previously reported by Zalinge \textit{et al.} \cite{Zalinge03} They used a laser spot as a heater to introduce $\boldsymbol{\nabla} T$, and therefore the measured thermovoltages were primarily attributed to the phonon-drag.

The diffusion contribution can be selectively measured by introducing the gradient only to the electron temperature $T_\text{e}$, leaving the lattice temperature $T_\text{L}$ intact. This can be achieved by directly passing a moderately large current to a section of 2DEG designated to serve as a heater. \cite{Maximov04,Fujita10E} 
It will be practically very difficult, however, to apply this Joule heating technique to a Corbino device, since it requires the heating current to pass only along the outer (or inner) rim of the annularly patterned 2DEG.
In the present study, we take an alternative approach, microwave irradiation, for the electron heating. The microwaves propagating through a coplanar wave guide (CPW) \cite{Engel93} deposited on the surface of a 2DEG wafer are partially absorbed by, and raise the temperature of, the 2DEG underneath the slots of the CPW \cite{Schliewe01,Endo13K}. By placing a CPW along the outer periphery of a Corbino disk, we introduce a radial gradient in $T_\text{e}$, and measure the resulting thermovoltages $V_{rr}$. We observe the behavior of $V_{rr}$ expected for the diffusion contribution in the QH systems in the Corbino geometry.

\begin{figure}
\begin{center}
\includegraphics[bbllx=60,bblly=30,bburx=790,bbury=510,width=8.6cm]{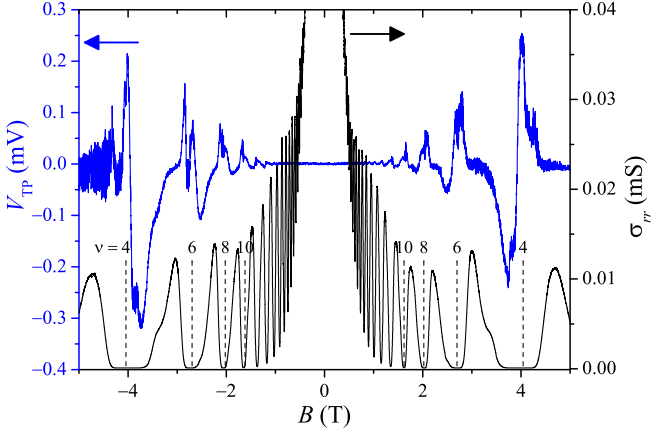}
\end{center}
\caption{(Color online) Radial thermovoltage $V_\text{TP} = -V_{rr}$ (left axis) and conductivity $\sigma_{rr}$ (right axis) measured in the Corbino device at $T_\text{bath} =1.4$ K\@. Positions of (even) integer fillings are indicated by vertical dashed lines.}
\label{VTPsigma}
\end{figure}

First, we describe the difference between the thermopower measured in the Hall bar geometry and that measured in the Corbino setup \cite{Zalinge03,Barlas12}. We start with the transport equation that relates the current density $\boldsymbol{j}$ to the applied electric field $\boldsymbol{E}$ and the temperature gradient $\boldsymbol{\nabla} T$ \cite{Fletcher99},%
\begin{equation}
\boldsymbol{j} = \hat{\sigma} \boldsymbol{E} - \hat{\epsilon} \boldsymbol{\nabla} T,
\label{CurrDens}
\end{equation}
where $\hat{\sigma}$ and $\hat{\epsilon}=\hat{\epsilon}^\text{(d)}+\hat{\epsilon}^\text{(g)}$ represent the conductivity tensor and the thermoelectric conductivity tensor, respectively. Here and in what follows, we use the superfixes (d) and (g) to denote the diffusion and the phonon-drag contributions, respectively.
The thermopower tensor $\hat{S}$ is defined by the relation $\boldsymbol{E} = \hat{S} \boldsymbol{\nabla} T$. In the Hall bar geometry, $\hat{S}$ is measured under the condition that no current flows, namely, $\boldsymbol{j} = 0$ in eq.\ (\ref{CurrDens}), and thus $\hat{S} = \hat{\sigma}^{-1}\hat{\epsilon}$. For an isotropic 2DEG ($\sigma_{xx} = \sigma_{yy}$ and $\sigma_{xy} = -\sigma_{yx}$), the longitudinal (Seebeck) and the transverse (Nernst) components are written, noting that $\sigma_{xx} \ll |\sigma_{yx}|$ in a strong magnetic field, as,
\begin{equation}
S_{xx}^\text{H} =  \frac{\sigma_{xx} \epsilon_{xx} + \sigma_{yx} \epsilon_{yx}}{\sigma_{xx}^2+\sigma_{yx}^2} \simeq \frac{\epsilon_{yx}}{\sigma_{yx}}
\label{Sxx}
\end{equation}
and
\begin{equation}
S_{yx}^\text{H} = \frac{\sigma_{xx} \epsilon_{yx} - \sigma_{yx} \epsilon_{xx}}{\sigma_{xx}^2+\sigma_{yx}^2} \simeq -\frac{\epsilon_{xx}}{\sigma_{yx}},
\label{Syx}
\end{equation}
respectively.
In the Corbino geometry, on the other hand, the constraint on the current is imposed only on the radial direction $j_r = 0$. With the temperature gradient $\nabla_r T$ applied in the radial direction, and noting that the electric field can have the component only in the radial direction $E_r$, we have $j_{r} = \sigma_{rr} E_r - \epsilon_{rr} \nabla_r T =0$, leading to the thermopower,
\begin{equation}
S_{rr}^\text{C} = \frac{E_r}{\nabla_r T} = \frac{\epsilon_{rr}}{\sigma_{rr}}.
\label{Srr}
\end{equation}
The diffusion contribution to the thermoelectric conductivity tensor is related to the conductivity tensor as \cite{Fletcher99}
\begin{equation}
\hat{\epsilon}^\text{(d)}(T) = -\frac{1}{eT} \int_0^\infty d\varepsilon \left( -\frac{\partial f}{\partial \varepsilon} \right) (\varepsilon-\zeta) \hat{\sigma}_0(\varepsilon ),
\label{epsdT}
\end{equation}
where $f(\varepsilon)=\{ 1+\exp[(\varepsilon - \zeta)/(k_\text{B} T)] \}^{-1}$ is the Fermi-Dirac distribution function, $\zeta$ the chemical potential, and $\hat{\sigma}_0(\varepsilon )$ the conductivity tensor at $T =0$ with
\begin{equation}
\hat{\sigma}(T) = \int_0^\infty d\varepsilon \left( -\frac{\partial f}{\partial \varepsilon} \right) \hat{\sigma}_0(\varepsilon ).
\label{sgmT}
\end{equation}
The thermopower tensor can also be decomposed into the two contributions, $\hat{S}^\text{(d)} = \hat{\sigma}^{-1}\hat{\epsilon}^\text{(d)}$ and $\hat{S}^\text{(g)} = \hat{\sigma}^{-1}\hat{\epsilon}^\text{(g)}$. From eqs.\ (\ref{Sxx})-(\ref{sgmT}) and the well-known behavior of $\hat{\sigma}$ in the QH systems \cite{PrangeGirvin90}, it can be seen that the diffusion contributions measured in the Hall bar geometry $S_{xx}^\text{H(d)}$ and $S_{yx}^\text{H(d)}$ both vanish in the QH plateaus \cite{Jonson84}. In between two consecutive QH states, $S_{xx}^\text{H(d)}$ takes a negative peak, while $S_{yx}^\text{H(d)}$ alternates sign, taking positive, zero, then negative values with the increase of the filling factor $\nu$ (or the decrease of $B$) for $B > 0$, with $S_{xx}^\text{H(d)}(-B) = S_{xx}^\text{H(d)}(B)$ and $S_{yx}^\text{H(d)}(-B) = -S_{yx}^\text{H(d)}(B)$ \cite{Jonson84}.
The diffusion Corbino thermopower $S_{rr}^\text{C(d)}$ resembles $S_{yx}^\text{H(d)}$ in that both are sensitive to the diagonal component of $\hat{\epsilon}^\text{(d)}$, and therefore similar behaviors are expected in the inter-QH regime, except that $S_{rr}^\text{C(d)}$ is symmetric under the inversion of the magnetic field: $S_{rr}^\text{C(d)}(-B) = S_{rr}^\text{C(d)}(B)$. Importantly, however, they are completely different in the QH states, where the behavior of $S_{rr}^\text{C(d)}$ cannot be readily discerned from eq.\ (\ref{Srr}) since both the numerator $\epsilon_{rr}^\text{(d)}$ and the denominator $\sigma_{rr}$ vanish with $T \rightarrow 0$. 
\begin{figure}
\begin{center}
\includegraphics[bbllx=20,bblly=165,bburx=530,bbury=770,width=8.6cm]{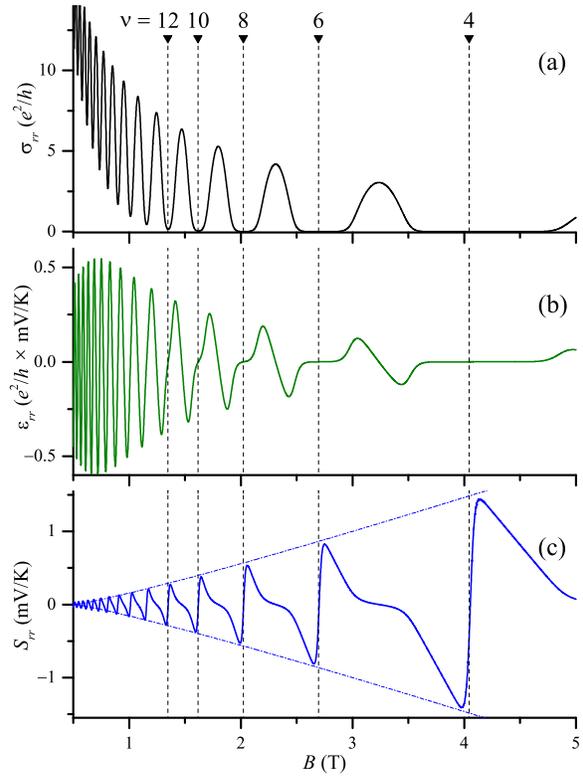}
\end{center}
\caption{(Color online) Transport coefficients at 1.5 K in the Corbino geometry calculated using SCBA, assuming complete spin degeneracy, $\Gamma$(meV) $ = 0.63 \sqrt{B(\text{T})}$, $n_\text{e} = 3.9\times 10^{15}$ m$^{-2}$, and $\zeta = 14$ meV\@. (a) Conductivity $\sigma_{rr}$. (b) Diffusion thermoelectric conductivity $\epsilon_{rr}^\text{(d)}$. (c) Diffusion thermopower $S_{rr}^\text{C(d)} = \epsilon_{rr}^\text{(d)} / \sigma_{rr}$. The dot-dashed lines in (c) represent $\pm (\hbar \omega_\text{c}/2 - \Gamma)/(e T)$. Positions of (even) integer fillings are indicated by vertical dashed lines.}
\label{SCBAcalc}
\end{figure}
A theoretical calculation \cite{Barlas12} that treats disorder within the self-consistent Born approximation (SCBA) reveals that $S_{rr}^\text{C(d)}$ takes large but finite values $\sim \pm (\hbar \omega_\text{c}/2 - \Gamma)/(e T)$ near the center of the QH plateaus, taking the positive (negative) sign for the filling below (above) an integer value and changing sign at the plateau center \cite{NoteSign}. Here, $\hbar \omega_\text{c}$ and $\Gamma$ represent the cyclotron energy and the disorder broadened Landau-level (LL) width, respectively.
Thus the sign alternation takes place both at the midpoint between and at the center of the QH plateaus, and $S_{rr}^\text{C(d)} < 0$ ($S_{rr}^\text{C(d)} > 0$) for the fillings at which the transport is dominated by the electrons (holes).
In Fig.\ \ref{SCBAcalc}, we plot $\sigma_{rr}$, $\epsilon_{rr}^\text{(d)}$, and $S_{rr}^\text{C(d)} = \epsilon_{rr}^\text{(d)} / \sigma_{rr}$ as functions of $B$, calculated by replacing into eqs.\ (\ref{sgmT}) and (\ref{epsdT}) the zero-temperature longitudinal conductivity in SCBA \cite{AndoRev82}
\begin{equation}
\sigma_{0,rr}(\varepsilon)=\frac{e^2}{h} \frac{4}{\pi} \sum\limits_N \left( N+\frac{1}{2} \right) \mathrm{Max} \left\{ 1-\left( \frac{\varepsilon - \varepsilon_N}{\Gamma} \right)^2 ,0 \right\},
\end{equation}
with $\varepsilon_N = (N+1/2) \hbar \omega_\text{c}$. In the calculation, we used a fixed value $\zeta = 14$ meV (corresponding to $n_\text{e} = 3.9 \times 10^{15}$ m$^{-2}$ at $B = 0$) \cite{fixedCP}, assumed that the spin degeneracy is not resolved, and set the width $\Gamma \propto \sqrt{B}$ to roughly reproduce the width of inter-QH peaks in $\sigma_{rr}$ observed experimentally (Fig.\ \ref{VTPsigma}).

The Corbino thermopower reported previously \cite{Zalinge03} exhibited the magnetic-field dependence distinctly different from that shown in Fig.\ \ref{SCBAcalc}(c): peaks were observed at the QH states without the sign alternation. As mentioned earlier, their heating technique sets up the temperature gradient both in $T_\text{e}$ and $T_\text{L}$, and thus their thermopowers were attributed to the phonon drag \cite{Zalinge03}. The diagonal component  $\epsilon_{xx}^\text{(g)}$ is theoretically shown to be absent \cite{Fromhold93}, and accordingly $S_{rr}^\text{C(g)}$ should also vanish, in a uniform and isotropic 2DEG\@.  The observed $S_{rr}^\text{C(g)}$ was interpreted \cite{Zalinge03} in terms of inhomogeneity \cite{Tieke97} or anisotropy \cite{Butcher98} inevitably present in a real 2DEG\@.
 
%
\begin{figure}
\begin{center}
\includegraphics[bbllx=50,bblly=70,bburx=800,bbury=530,width=8.6cm]{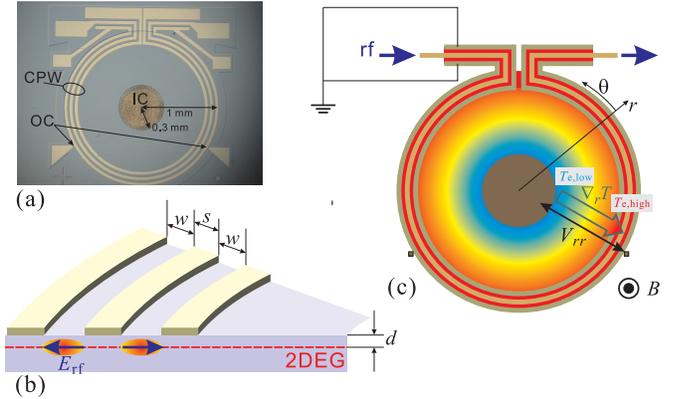}
\end{center}
\caption{(Color online) (a) Optical micrograph of the Corbino device. CPW: coplanar waveguide. IC: inner contact. OC: outer contact. (b) Close-up cross section of the vicinity of CPW; $w = 28$ $\mu$m, $s = 40$ $\mu$m, and $d = 60$ nm. 2DEG beneath the slots of CPW is heated by the rf electric field $E_\text{rf}$. (c) Schematic drawing of the Corbino device. Radial temperature gradient $\nabla_r T$ is introduced by the microwave heating, resulting in radial thermovoltage $V_{rr}$ between IC and OC\@.}
\label{sample}
\end{figure}
The Corbino device employed in the present study is schematically depicted in Fig.\ \ref{sample}. The device is prepared from a conventional GaAs/AlGaAs wafer with the 2DEG residing at the depth $d = 60$ nm from the surface. A disk mesa with the radius $R_\text{M} = 1$ mm is defined by electron-beam (EB) lithography and wet etching. A metallic (Au/Ti) CPW composed of a center electrode flanked by two grounded electrodes (all having the width $s = 40$ $\mu$m), separated by slots with the width $w = 28$ $\mu$m, is installed along the outer rim of the mesa using EB lithography. The CPW is designed to have the characteristic impedance $Z_0 =50$ $\Omega$. Ohmic contacts (Au-Ge-Ni alloy) are made at the center (inner contact, IC) and at the outer edge (outer contact, OC) of the disk. Since the device is immersed in the liquid helium during the measurement, the 2DEG is expected to be cooled down toward the bath temperature $T_\text{bath}$ through the contacts. The IC is designed to have a large radius $R_\text{IC} = 0.3$ mm so as to serve also as the low-temperature anchor of the device, whereas the OCs, located on short (23 $\mu$m) 2DEG arms protruding from the disk, are made small ($8 \times 10$ $\mu$m$^2$) to minimize the disturbance to the temperature gradient.  

Microwaves with the frequency 500 MHz are injected into the CPW from one end using a network analyzer (Agilent E5062). The microwaves capacitively couple to the 2DEG and are absorbed by the 2DEG beneath the slots, heating the electrons, thereby introducing the electron-temperature gradient $\boldsymbol{\nabla} T_\text{e}$ toward the IC\@. The resulting thermovoltages are measured as the dc voltages between IC and OC, $V_\text{TP} = -V_{rr} = V_\text{IC}-V_\text{OC}$. Here, the sign of $V_\text{TP}$ is defined so that the electric field $\boldsymbol{E}$ is parallel (anti-parallel) to $\boldsymbol{\nabla} T_\text{e}$ when $V_\text{TP} > 0$ ($V_\text{TP} < 0$) and thus the sign of $V_\text{TP}$ coincides with that of $S_{rr}$. By detecting the transmission of the microwaves through the CPW with the network analyzer, one can also measure the longitudinal conductivity $\sigma_{\theta \theta} = \sigma_{xx}$ of the 2DEG underneath the slots \cite{Engel93,Schliewe01,Endo13K}. From the amplitudes of the Shubnikov-de Haas (SdH) oscillations in the measured $\sigma_{\theta \theta}$, one can, in turn, estimate $T_\text{e}$ of the electrons heated by the microwaves \cite{Schliewe01,Endo13K}, which can be used to translate $V_\text{TP}$ into $S_{rr}$ (see the discussion below).
Measurements are performed at $T_\text{bath} = 1.4$ K in a $^4$He cryostat using a probe equipped with semi-rigid coaxial cables.

%
The main achievement in the present study is displayed in Fig.\ \ref{VTPsigma}, which plots the measured radial thermovoltages $V_\text{TP}$ along with the (radial) longitudinal conductivity $\sigma_{rr}$. The latter is acquired by the same set of the contacts, IC and OC, as the former; here, $\sigma_{rr}$ is conveniently obtained from the two-terminal resistance $r_2$ measured with a small current 0.1 nA by a standard low-frequency ac (13 Hz) lock-in technique after the subtraction of the series resistance $r_\text{s}$, $\sigma_{rr} = (2 \pi)^{-1} (r_2-r_\text{s})^{-1} \ln (R_\text{M}/R_\text{IC})$, which suffices for the purpose of identifying the positions of the QH plateaus.
It can be seen that the spin degeneracy remains unresolved in the magnetic-field range and the temperature in the present study. The electron density $n_\text{e}$ deduced from the frequency of the SdH oscillations and the mobility $\mu$ inferred from the low-field semiclassical background $\propto [1+(\mu B)^2]^{-1}$ of $\sigma_{rr}$ are $n_\text{e} = 3.9 \times 10^{15}$ m$^{-2}$ and $\mu \sim 10$ m$^2$V$^{-1}$s$^{-1}$, respectively. 

The measured $V_\text{TP}$ qualitatively exhibits the behavior expected for the Corbino thermopower described 
above [see Fig.\ \ref{SCBAcalc}(c) and note that $S_{rr}^\text{C(d)}(-B) = S_{rr}^\text{C(d)}(B)$]. Notably, within the QH plateau regime where $\sigma_{rr} = 0$, $V_\text{TP}$ takes large positive or negative values, decreases with increasing $\nu$ and crosses zero, inverting the sign, at the center of the plateaus (positions of exact even integer fillings calculated from $n_\text{e}$) indicated by the vertical dashed lines. In the region between the QH states, where the chemical potential $\zeta$ lies in the extended states near the center of Landau levels (LLs), the measured lineshape appears to have the longer span of flat regions compared to the calculated one. The discrepancy in the lineshape is possibly the outcome of the SCBA used in the calculation: it is well known that the semi-elliptical lineshape of the disorder-broadened LLs in the SCBA \cite{AndoRev82,Jonson84,Barlas12} does not reproduce the profile of the experimentally observed LLs, which is better described by either the Gaussian or the Lorentzian (see, e.g., ref.\ \citen{Zhu03}). In the QH plateau regime, on the other hand, $\zeta$ is located at the localized states and the transport at finite temperatures is carried out by electrons or holes thermally activated to the upper-lying extended states. Such situations are properly captured in SCBA, in which $\hat{\sigma}_0(\zeta) = 0$ when $\zeta$ is at the localized states and thus both $\hat{\epsilon}^{(\text{d})}(T)$ and $\hat{\sigma}(T)$ in eqs.\ (\ref{epsdT}) and (\ref{sgmT}) derive solely from the tails of $-\partial f/\partial \varepsilon$ where $\hat{\sigma}_0(\varepsilon)$ possesses a finite value.

\begin{figure}
\begin{center}
\includegraphics[bbllx=30,bblly=185,bburx=520,bbury=740,width=8.6cm]{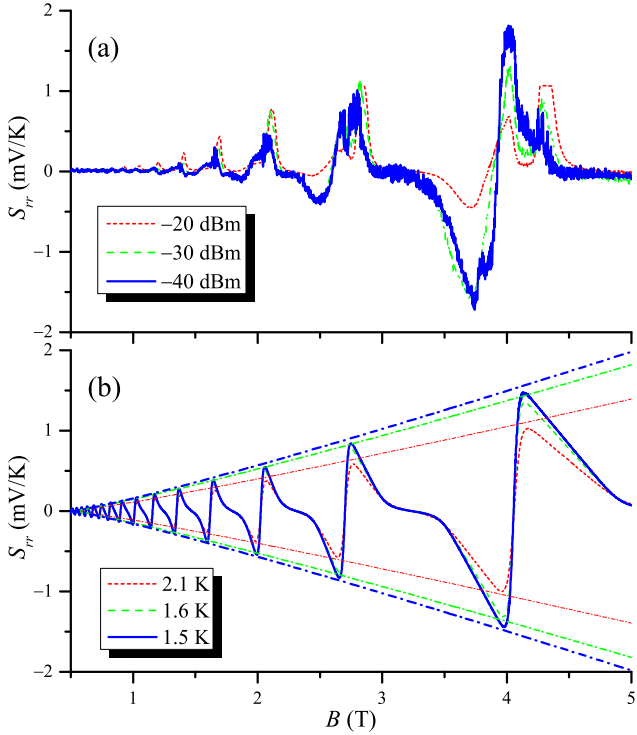}
\end{center}
\caption{(Color online) (a) Thermopowers $S_{rr}$ deduced from thermovoltages $V_\text{TP}$ measured at $T_\text{bath} =1.4$ K with three different microwave powers, $P_\text{NA} = -20$, $-30$, and $-40$ dBm, corresponding to $T_\text{e,high} \simeq 2.8$, 1.8, 1.5 K, respectively. (b) Thermopowers calculated using SCBA for three different temperatures, corresponding to $T_\text{ave} = (T_\text{e,high} + T_\text{bath})/2$ with $T_\text{e,high}$ in (a). The dot-dashed lines represent $\pm (\hbar \omega_\text{c}/2 - \Gamma)/(e T_\text{ave})$. }
\label{Spws}
\end{figure}
To be more quantitative, we make an attempt to translate $V_\text{TP}$ into $S_{rr}^\text{C(d)}$. To this end, we estimate the $T_\text{e,high}$ of the electrons 
heated by the microwaves, using the SdH amplitudes of $\sigma_{\theta \theta}$ deduced from the microwave transmission through the CPW \cite{Schliewe01,Endo13K}. We assume that $T_\text{e,low}$ at the IC equals the helium bath temperature $T_\text{bath} = T_\text{L} = 1.4$ K\@. With the temperature difference $\Delta T_\text{e} = T_\text{e,high}-T_\text{bath} = \int_{R_\text{IC}}^{R_\text{M}} dr \nabla_r T_\text{e}$, the thermovoltage $V_\text{TP} = \int_{R_\text{IC}}^{R_\text{M}} dr E_{rr}$ (note the definition of the sign of $V_\text{TP}$ mentioned above), and the relation $E_{rr} = S_{rr}^\text{C(d)} \nabla_r T_\text{e}$, we have $S_{rr}^\text{C(d)} = V_\text{TP}/\Delta T_\text{e}$ if we neglect the temperature dependence of $ S_{rr}^\text{C(d)}$ in the range $\Delta T_\text{e}$.  We made the measurements for three different microwave powers: the output powers of the network analyzer $P_\text{NA} = -40$, $-30$, and $-20$ dBm, corresponding to the powers at the entrance of the CPW (subtracting the power absorbed along the cables) 0.015, 0.15, and 1.5 $\mu$W, respectively. The $V_\text{TP}$ plotted in Fig.\ \ref{VTPsigma} were taken with the lowest power $P_\text{NA} = -40$ dBm. The estimated temperatures are $T_\text{e,high} = $ 1.5, 1.8, and 2.8 K, respectively, and the traces of $S_{rr}^\text{C(d)}$ obtained using these temperatures are plotted in Fig.\ \ref{Spws}(a). We can see that the oscillation amplitudes are in fairly good agreement with the traces calculated by the SCBA for $T = T_\text{ave} = (T_\text{e,high}+T_\text{bath})/2$ plotted in Fig.\ \ref{Spws}(b). Note, however, that the oscillation amplitudes of the calculated traces are of limited reliability owing to the use of rather uncertain values of $\Gamma$ in the calculation.

We found that the electron density $n_\text{e} = 4.1\times10^{15}$ m$^{-2}$ deduced from the microwave transmission ($\sigma_{\theta \theta}$) was slightly larger than that obtained from the two-terminal measurement ($\sigma_{rr}$), possibly signaling the  inhomogeneity, or the effect of the microwave irradiation and/or the presence of metallic CPW on the electron density. Consequently, the QH plateaus in $\sigma_{\theta \theta}$ take place at slightly higher magnetic fields compared to those in  $\sigma_{rr}$. Small structures corresponding to the center of the QH plateaus in $\sigma_{\theta \theta}$ can be identified in the $V_\text{TP}$ in Fig.\ \ref{VTPsigma} as a shoulder (e.g., at $\sim \pm 4.3$ T) or peak splitting (e.g., at $\nu \sim 6$). Features belonging to this second phase ascribable to the 2DEGs in the CPW slot regions increase their relative intensity with the increase of the microwave power, as can be seen in Fig.\ \ref{Spws}(a) \cite{OverR}. In addition, $V_\text{TP}$ measured with higher microwave powers contains positive background which increases with the magnetic field [see Fig.\ \ref{Spws}(a)], whose origin remains unidentified but could possibly be due to the intervention of the phonon-drag contribution; with an excessively high microwave power, the lattice can also be heated via electron-phonon interaction, leading to the emergence of the phonon-drag effect. In order to avoid these complications, and noting that $\hat{S}$ is defined by $\boldsymbol{E} = \hat{S} \boldsymbol{\nabla} T$ in the limit of vanishingly small $|\boldsymbol{\nabla} T|$, the microwave power should be kept as small as possible to obtain accurate values of $\hat{S}$.

To conclude, we have measured the diffusion contribution to the Corbino thermopower, employing microwave heating technique to introduce the electron-temperature gradient. Large values within the QH plateaus, with the sign reversal at the center of the plateaus, are observed, as predicted in a recent theory \cite{Barlas12}. The theory suggests that the non-Abelian nature of the quasiparticles in the $\nu = 5/2$ fractional QH state can be probed more efficiently in the Corbino geometry \cite{Barlas12} than in the more conventional Hall-bar geometry \cite{Yang09}. The experimental method in the present study, when applied to 2DEGs with the quality high enough to accommodate the $\nu = 5/2$ fractional QH state, can be an effective tool to investigate the exotic statistics of this intriguing fractional QH state.

\begin{acknowledgment}
This work was supported in part by Grant-in-Aid for Scientific Research (A) No.\ 18204029, (B) No.\ 20340101, and (C) No.\ 22560004 from Japan Society for the Promotion of Science.
\end{acknowledgment}


\bibliography{thermo,qhe,twodeg,wc,ourpps,NoteKbykCorbino}

\begin{thebibliography}{10}
\bibitem{NoteSign}
Note that the sign of the Corbino thermopower $Q_\text{C}$ in Ref.\
  \citen{Barlas12} is defined to be opposite to that of $S_{rr}^\text{C(d)}$ in
  the present study.
\bibitem{fixedCP}
Here, we neglected the possible dependence of $\zeta$ on $B$ for simplicity. In
  this approximation, $\zeta$ is allowed to be located for finite
  magnetic-field spans in the regions where the density of states totally
  vanishes in SCBA, which, in effect, represent the localized states.
\end{thebibliography}

\end{document}